\documentclass{iopart}
\usepackage{graphicx} % Include figure files
\usepackage{cite}
\usepackage{float}
\usepackage{color}
\begin{document}

\title[Direct measurement of the anisotropy field in Mn--Co--Ga]{Direct measurement of the magnetic anisotropy field in Mn--Ga and Mn--Co--Ga Heusler films}

\author{Ciar{\'a}n~Fowley\textsuperscript{1},
Siham~Ouardi\textsuperscript{2}, 
Takahide~Kubota\textsuperscript{3},
Yildirim~Oguz\textsuperscript{1}, 
Andreas~Neudert\textsuperscript{1}, 
Kilian~Lenz\textsuperscript{1},
Volker~Sluka\textsuperscript{1},
J{\"u}rgen~Lindner\textsuperscript{1},
Joseph~M.~Law\textsuperscript{4}, 
Shigemi~Mizukami\textsuperscript{3},
Gerhard~H.~Fecher\textsuperscript{2},
Claudia~Felser\textsuperscript{2} and 
Alina~M.~Deac\textsuperscript{1}}
% change order to fowley, ouardi     ... done ghf
% japanese group for sample growth?  ... done ghf
\address{\textsuperscript{1} Institute of Ion Beam Physics and Materials Research. Helmholtz-Zentrum Dresden-Rossendorf, Bautzner Landstrasse 400, 01328 Dresden, Germany.}
\address{\textsuperscript{2} Max-Planck-Institut f{\"u}r Chemische Physik fester Stoffe, N{\"o}thnitzer Str. 40, 01187 Dresden.}
\address{\textsuperscript{3} WPI-Advanced Institute for Materials Research (WPI-AIMR), Tohoku University, Sendai 980-8577, Japan}
\address{\textsuperscript{4} High Magnetic Field Laboratory, Helmholtz-Zentrum Dresden-Rossendorf, Bautzner Landstrasse 400, 01328 Dresden, Germany.}

\ead{c.fowley@hzdr.de}

%%%%%%%%%%%%%%%%%%%%%%%%%%%%%%%%%%%%%%%%%%%%%%%%%%%%%%%%%%%%%%%%%%%%%%%%%
\begin{abstract}

The static and dynamic magnetic properties of tetragonally distorted Mn--Ga based alloys 
were investigated. Static properties are determined in magnetic fields 
up to 6.5~T using SQUID magnetometry. 
For the pure Mn$_{1.6}$Ga film, the saturation magnetisation is 0.36~MA/m and 
the coercivity is 0.29~T. Partial substitution of Mn by Co results in 
Mn$_{2.6}$Co$_{0.3}$Ga$_{1.1}$. The saturation magnetisation of those films 
drops to 0.2~MA/m and the coercivity is increased to 1~T.

Time-resolved magneto-optical Kerr effect (TR-MOKE) is used to probe the high-frequency 
dynamics of Mn--Ga. The ferromagnetic resonance frequency extrapolated 
to zero-field is found to be 125~GHz with a Gilbert damping, $\alpha$, 
of 0.019. The anisotropy field is determined from both SQUID and TR-MOKE to be 
4.5~T, corresponding to an effective anisotropy density of 0.81~MJ/m$^3$.

Given the large anisotropy field of the Mn$_{2.6}$Co$_{0.3}$Ga$_{1.1}$ film, 
pulsed magnetic fields up to 60~T are used to determine the field strength 
required to saturate the film in the plane. For this, the extraordinary Hall 
effect was employed as a probe of the local magnetisation. By integrating the 
reconstructed in--plane magnetisation curve, the effective anisotropy energy 
density for Mn$_{2.6}$Co$_{0.3}$Ga$_{1.1}$ is determined to be 1.23~MJ/m$^3$.

\end{abstract}

%Uncomment for PACS numbers title message
%\pacs{00.00, 20.00, 42.10}
% Keywords required only for MST, PB, PMB, PM, JOA, JOB? 
%\vspace{2pc}
%\noindent{\it Keywords}: Article preparation, IOP journals
% Uncomment for Submitted to journal title message
\submitto{\JPD}
% Comment out if separate title page not required
\maketitle

%%%%%%%%%%%%%%%%%%%%%%%%%%%%%%%%%%%%%%%%%%%%%%%%%%%%%%%%%%%%%%%%%%%%%%%%%%%%%%%%
\section{Introduction}

There has been a recent resurge in research on Heusler alloys, in particular Mn-based 
ferrimagnetic compounds, for both magnetic storage and spin-transfer-torque 
applications due to the ability to widely tune the magnetic properties with 
varying composition~\cite{BFW07,AWF11,KRV11,OKF12,KOM13,RBB13}. The addition of Co to Mn-Ga allows the subtle tuning of magnetic properties, such as uniaxial anisotropy and saturation magnetisation~\cite{OKF12}.
These films can possess very high uniaxial anisotropy~\cite{MWS11,KRS14}. They are 
promising for future rare-earth free permanent magnets~\cite{Co14}; highly 
stable magnetic recording elements scalable down to 10~nm bit size~\cite{KRV11}; 
spin-polarised electrodes for tunnel magnetoresistance devices~\cite{KMW11,MKM12}; as 
well as active elements in next generation spin-transfer torque devices such as THz-band spin transfer oscillators, due to their high ferromagnetic resonance frequencies 
and low Gilbert damping~\cite{MWS11}.

In these films, anisotropy fields of several tens of Tesla's have been 
reported~\cite{OKF12,KRV11,RBB13}. Combined with their low magnetisation, 
typically below 0.5~MA/m, traditional SQUID magnetometry and conventional 
ferromagnetic resonance (FMR) techniques are not so well suited for magnetic 
characterisation.
For the majority of recent reports the anisotropy 
field is extrapolated from the intersection of the low field linear slope of the hard-axis magnetisation curve to the easy-axis saturation magnetisation. This intersection occurs at a field which is, in general, 
beyond the machine limit. For SQUID magnetometry this requires careful subtraction of the diamagnetic background signal \cite{KKE13}.
It should also be pointed out that most studies to 
date have also been performed on thick samples, while more technologically 
relevant thinner films have shown reduced anisotropies \cite{KKE13}.  
Finally, measurement responses to standard SQUID- and FMR-based studies are also directly proportional to the total magnetic moment present. 
An alternative approach is to exploit 
electrical or optical detection techniques to characterise such materials. 
The direct measurement of the anisotropy field is especially required if there are additional, for example in-plane, anisotropy components to be considered in the material under investigation.

Time-resolved magneto-optical Kerr effect (TR-MOKE) has already been 
successfully used to characterise the high frequency dynamics of Mn--Ga thin 
films, showing precession frequencies between 200~GHz and 300~GHz~\cite{MWS11}. 
The frequency of oscillation as a function of applied field can be fit with the Kittel formula.
Such measurements therefore yield not only the ferromagentic resonance (FMR) frequency and the Gilbert damping parameter, $\alpha$, but also the effective anisotropy field and energy density ($\mu_0H_k$ and $K_{eff}$, respectively). 
Provided the saturation magnetisation is known, the intrinsic 
uniaxial anisotropy $K_u$ can be determined from $K_{eff}=K_u-\frac{1}{2}\mu_0 M^2_S$, 
where $\frac{1}{2}\mu_0 M^2_S$ represents the demagnetising energy of an extended 
thin-film.

The extraordinary Hall effect (EHE) is a very useful characterisation tool for these
perpendicularly magnetised materials~\cite{GEI13}. When a current is applied along a certain direction in a film, the transverse resistivity ($\rho_{xy}$) is directly proportional to the out--of--plane magnetisation component ($M_z$) via the extraordinary Hall coefficient, $R_{EHE}$\cite{NSO10}.
For perpendicular anisotropy materials, if we apply an external field along the easy axis of the film and switch the magnetisation, and therefore $\rho_{xy}$, we will obtain an electrical equivalent of the magnetic hysteresis loop.
Similarly, when the field is applied along the hard axis, the saturation of the magnetisation can be seen as a gradual change and saturation of $\rho_{xy}$. 
From the hard axis data, the anisotropy field, $\mu_0H_k$, can be determined.
EHE allows for characterisation of these high-anisotropy 
films beyond what is normally accessible from standard magnetometry for several 
reasons: firstly, it is a transport technique and not a magnetometry technique, 
which means it can be used to measure the magnetic response of volumes of 
material and/or patterned structures which would be otherwise undetectable; secondly, an inherent advantage is that $\rho_{xy}$ exhibits an inverse thickness dependence meaning that the EHE signal 
is larger for thinner films; thirdly, $R_{EHE}$ in ferromagnetic materials is 
large meaning that even though $M_S$ may be low, $\rho_{xy}$ can be high. 
These advantages make EHE an ideal probe of the magnetisation at lower 
thickness and/or confined geometry. 

In the search for materials with higher perpendicular anisotropy and lower saturation magnetisation for spin-transfer-torque applications, alternative characterisation techniques, such as those outlined above, will become increasingly more important. To exemplify this, we investigate \emph{L}1$_0$ Mn$_{1.6}$Ga (Mn--Ga) and Mn$_{2.6}$Co$_{0.3}$Ga$_{1.1}$ (Mn--Co--Ga) using TR-MOKE and EHE in high magnetic fields. Both materials posses high uniaxial anisotropy. In particular, the chosen composition of Mn--Co--Ga has been shown to exhibit very low saturation magnetisation while retaining high anisotropy \cite{OKF12}. They therefore represent ideal samples for determining the usefulness of the techniques as the properties are favourable for applications while at the same time may prove difficult to determine with more traditional methods, especially at reduced thicknesses~\cite{GEI13}. We show that, in particular, EHE at high magnetic fields is an ideal method for determining the anisotropy field of such Heusler systems.

%%%%%%%%%%%%%%%%%%%%%%%%%%%%%%%%%%%%%%%%%%%%%%%%%%%%%%%%%%%%%%%%%%%%%%%%%%%%%%%%
\section{Experimental details}
%%%%%%%%%%%%%%%%%%%%%%%%%%%%%%%%%%%%%%%%%%%%%%%%%%%%%%%%%%
\begin{figure}[H]
\centering
   \includegraphics[width=9cm]{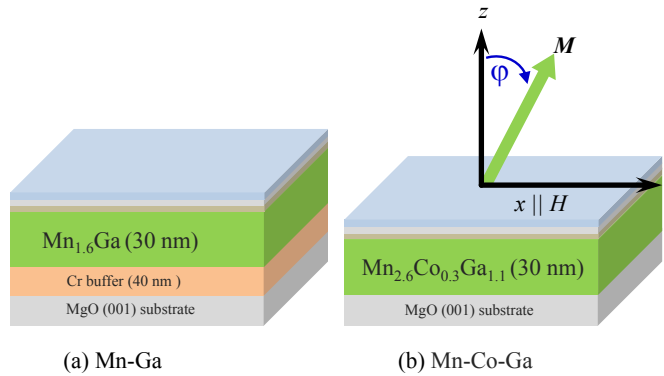}
   \caption{Sketch of the Mn$_{1.6}$Ga (a) and Mn$_{2.6}$Co$_{0.3}$Ga$_{1.1}$ (b) film stacks.
            The topmost layers are Mg, MgO, and AlO$_x$ to protect the sample from oxidation (see text for thicknesses).}
\label{fig:structure}
\end{figure}
%%%%%%%%%%%%%%%%%%%%%%%%%%%%%%%%%%%%%%%%%%%%%%%%%%%%%%%%%%%%%%%%%%%%
Tetragonal Mn--Ga and Mn--Co--Ga 
thin films were grown by ultra-high-vacuum sputtering on single crystal MgO(001) substrates. 
Specific details on sample fabrication and growth conditions can be found in 
Reference~\cite{OKF12}. The stacking structures of the multi-layered films are:\\ 
MgO(100) substrate~/ Cr(40)~/ Mn$_{1.6}$Ga(30)~/ Mg(0.4)~/ MgO(2.0)~/ AlO$_x$(1.3) and \\ 
MgO(100) substrate~/ Mn$_{2.6}$Co$_{0.3}$Ga$_{1.1}$(30)~/ Mg(0.4)~/ MgO(2.0)~/ AlO$_x$(1.3) \\ 
as sketched in Figures~\ref{fig:structure}(a)) and (b), respectively. The 
thickness, in nm, is appended in brackets after each material. The topmost 
three layers were added to simulate a tunnelling barrier and to protect the 
sample from oxidation.

Low-field magnetization data were obtained by magnetic field dependent magnetisation measurements at 300~K, up to an applied external field of 6.5~T using conventional SQUID magnetometry. TR-MOKE was 
used to characterise the high frequency dynamic properties of the 
Mn$_{1.6}$Ga film. An 800~nm pump beam with a power of 
31.2~mW at the sample, leads to ultra-fast demagnetisation of the 
sample. The laser pulses were 100~fs in length with a repetition rate of 5.2~MHz. A fixed-delay probe beam of 400~nm, with a power of 105~$\mu$W at the sample is then used to probe the excited dynamics via lock-in detection. The diameter of the spot sizes of the pump and probe beams were 17$~\mu$m and 5~$\mu$m, respectively. The pump and probe fluences, calculated from the incident power, beam diameter and repetition frequency 1.32~mJ/cm$^2$ and 0.05~mJ/cm$^2$, respectively.
Time resolution is obtained by a variable delay line on the pump beam which allows for the measurement of changes in magnetisation both
before and after the demagnetisation process. 

EHE was measured initially using a DC current of 1~mA at room temperature in a 
magnetotransport set-up capable of applying fields up to 1.6~T. Subsequent EHE 
measurements were performed in a cryostat at 77~K using pulsed magnetic fields 
at the High Magnetic Field Laboratory located in the Helmholtz-Zentrum 
Dresden-Rossendorf. The pulsed magnet which was used had a rise time of 7~ms, a fall time of 24~ms 
and a maximum field of approximately 60~T. An sinusoidal AC current of 1~mA was applied to the sample with a frequency of 
47.62~kHz. The transverse voltage was measured during the 
magnetic field pulse and the signal was locked-in post experiment via a digital 
lock-in program.

EHE scans along the easy and hard axes, in combination with the evaluated $M_S$ 
from SQUID, allows for the direct determination of the anisotropy field, 
$\mu_0H_k$, as well as the effective anisotropy energy, $K_{eff}$, by integration of 
normalised magnetisation loops.

%%%%%%%%%%%%%%%%%%%%%%%%%%%%%%%%%%%%%%%%%%%%%%%%%%%%%%%%%%%%%%%%%%%%%%%%%%%%%%%%
\section{Magnetic properties}

\begin{figure}[htb]
\centering
\includegraphics[width=10cm]{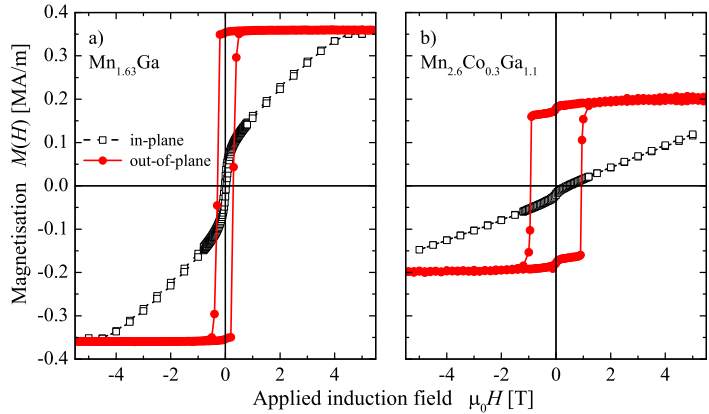}
\caption{ Static magnetic properties at 300~K for a) Mn--Ga and b) Mn--Co--Ga films.
          Both films exhibit strong perpendicular magnetic anisotropy, with approximately square hysteresis loops and $M_r/M_S\approx1$. 
          The addition of Co results in a drop of the magnetisation accompanied by an increase in both the 
          coercive field and the anisotropy field (increased beyond 5~T). }
\label{fig:squid}
\end{figure}

The initial magnetic characterisation of the thin films using SQUID is shown in 
Figure~\ref{fig:squid}. The pure Mn--Ga film -- shown in 
figure~\ref{fig:squid}~a) -- exhibits sharp hysteretic switching when the field 
is applied in the out--of--plane direction (closed red circles) with a coercive field, $\mu_0H_c$, 
of 290~mT. The data show clearly that the easy axis is along the film 
normal. A small in-plane component appears in the hard axis measurements consistent with an intrinsic canted moment on the 2\emph{b} Mn sub-lattice \cite{RBB13}. The saturation magnetisation, $M_S$, is 0.36~MA/m. In--plane (open black squares) magnetisation measurements give an anisotropy field, $\mu_0H_K$, of 4.5~T, which 
was also verified by vibrating sample magnetometry (VSM) up to 14~T (not shown here). 
These values correspond to an anisotropy energy density $K_{eff}$ of 
0.81~MJ/m$^3$.

%%%%%%%%%%%%%%%%%%%%%%%%%%%%%%%%%%%%%%%%%%%%%%%%%%%%%%%%%%%%%%%%%%%%%%%%%%%%%%%%
%%%%%%%%%%%%%%%%%%%%%%%%%%%%%%%%%%%%%%%%%%%%%%%%%%%%%%%%%%%%%%%%%%%%%%%%%%%%%%%%

The addition of Co into Mn--Ga, shown in figure~\ref{fig:squid}~b), leads to a 
reduction of the $M_S$ to 0.2~MA/m and an accompanied 
increase of $\mu_0H_c$ to 928~mT. 
The anisotropy field is 
also increased to a value beyond 5~T. For the in--plane curve, 
the same diamagnetic background as in the out--of--plane 
measurement was subtracted from the raw data, however as can be seen from the data, an accurate determination of the anisotropy field is not possible.
VSM magnetometry up to 14~T was not able to confirm the anisotropy field due to a low magnetisation signal (not shown).
We note that, as opposed to the Mn--Ga film we have a soft magnetic component in both the in-plane and out-of-plane magnetisation curves indicating that we do not have the same canted moment as in the pure film.
Rather, the data seem to indicate a segregated phase lacking any anisotropy axis. 
This could be initially attributed to Co clusters in the film or a lack of full epitaxial growth due to the lack of seed layer.
%we must therefore measure it directly using EHE in pulsed magnetic fields. 

The $M(H)$ hysteresis curves of both materials are nearly rectangular. The energy product is thus 
given by $E=B_r ~ \textrm{x}~ H_c$, where $B_r$ and $H_c$ are the remanence and the coercive field, respectively.
The out--of--plane energy products are 104~kJ/m$^3$ for Mn$_{1.6}$Ga and 165~kJ/m$^3$ for Mn$_{2.6}$Co$_{0.3}$Ga$_{1.1}$.
The addition of Co leads to a reduction of the maximum energy product, $BH_{\max}$, from about 40 to 10~kJ/m$^3$.
The values of the energy product and $BH_{\max}$ for Mn$_{2.6}$Co$_{0.3}$Ga$_{1.1}$ are of the same order as
bulk Mn$_3$Ga (cf. Reference~\cite{WBF08}).

%%%%%%%%%%%%%%%%%%%%%%%%%%%%%%%%%%%%%%%%%%%%%%%%%%%%%%%%%%%%%%%%%%%%%%%%%%%%%%%%
\begin{figure}[htb]
\centering
\includegraphics[width=8cm]{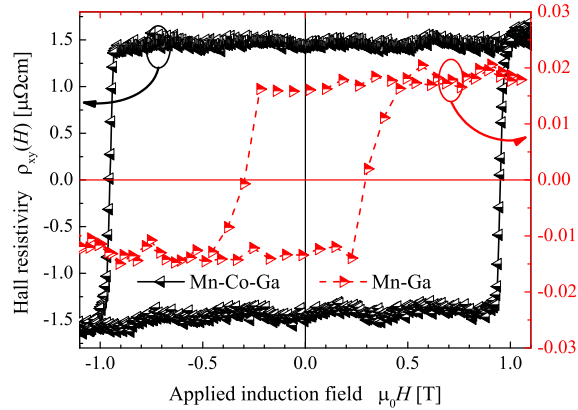}
\caption{ Extraordinary Hall effect data measured at room temperature for Mn--Ga and Mn--Co--Ga films. The coercive fields match exactly with those obtained from SQUID in figure~\ref{fig:squid}. 
          Note that the magnitudes of $\rho_{xy}$ for the two films differ by about two orders of magnitude.}
\label{fig:ehe}
\end{figure}
%%%%%%%%%%%%%%%%%%%%%%%%%%%%%%%%%%%%%%%%%%%%%%%%%%%%%%%%%%%%%%%%%%%%%%%%%%%%%%%%

Figure~\ref{fig:ehe} shows the EHE curves obtained from 5~mm $\times$ 5~mm square films measured in the van~der~Pauw geometry. For both films, 
the coercivity is identical to that obtained from SQUID measurements. The 
magnitude of $\rho_{xy}$ for Mn--Ga is much lower than that obtained for the 
Mn--Co--Ga film. Although the underlying Cr layer is responsible for current shunting, it is anyway expected that pure Mn--Ga films show a lower Hall signal with improving crystal quality \cite{GEI13}. 
It can be seen that the EHE curve for the Co--Mn--Ga films does not trace out the additional change in magnetisation close to zero field as seen in figure \ref{fig:squid} (b).
As has been calculated for Co-doped Mn--Ga films, Co substitutes Mn randomly at both the 2\emph{b} and 4\emph{d} positions, fills the minority band at \emph{E$_F$} and leads to the localisation of electrons in the minority band \cite{CKF13}.
Assuming that the soft magnetic component seen in SQUID is related solely to Co substitution, it is likely that, due to the increased electron localisation, the contribution of Co to the magnetotransport is diminished, therefore the same soft phase is not reproduced in the EHE measurement.

%%%%%%%%%%%%%%%%%%%%%%%%%%%%%%%%%%%%%%%%%%%%%%%%%%%%%%%%%%%%%%%%%%%%%%%%%%%%%%%%
\section{Time-resolved MOKE}

TR-MOKE was used to evaluate the effective anisotropy of the Mn--Ga film. A 
magnetic field of 0.65 to 1.3~T was applied at 60$^{\circ}$ to the film normal in order to cant the 
magnetisation away from the easy axis. This provides a projection of the 
precession along the film normal which is then measured in the polar MOKE 
geometry.

%%%%%%%%%%%%%%%%%%%%%%%%%%%%%%%%%%%%%%%%%%%%%%%%%%%%%%%%%%%%%%%%%%%%%%%%%%%%%%%%
\begin{figure}[htb]
\centering
\includegraphics[width=6cm]{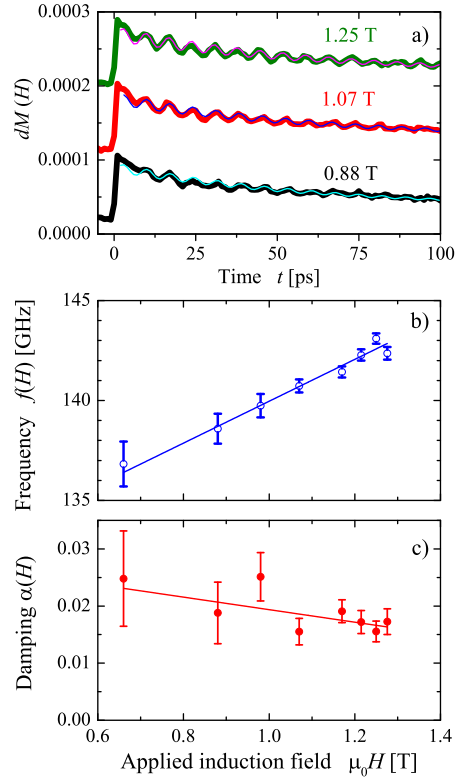}
\caption{ a) TR-MOKE spectra (bold lines) together with fitting curves (thin lines) for the Mn--Ga film at various fields applied $60^{\circ}$ to the film normal. 
          b) Frequency and c) damping as a function of applied field strength. 
          Lines in b) and c) result from linear fits. } 
\label{fig:mngatrmoke}
\end{figure}
%%%%%%%%%%%%%%%%%%%%%%%%%%%%%%%%%%%%%%%%%%%%%%%%%%%%%%%%%%%%%%%%%%%%%%%%%%%%%%%%

Figure~\ref{fig:mngatrmoke}~a) shows several TR-MOKE spectra at different applied magnetic fields as well as the fitted curves from which the frequency and damping are extracted. 
It can be seen in this 
figure that the magnetisation, after the initial demagnetisation pulse at $t=0$, 
oscillates around the effective field direction leading to a characteristic 
oscillation of the optical signal. The ferromagnetic resonance frequency 
($f_{res}$) and the Gilbert damping parameter ($\alpha$) are extracted by 
fitting these curves with the Kittel formula for the uniform mode. $f_{res}$ and 
$\alpha$ are plotted in figure~\ref{fig:mngatrmoke}~b) and~c), respectively. 
The precession frequency was found to vary between 136~GHz and 142~GHz in the aforementioned field range. 
The slope of the fitted line in figure \ref{fig:mngatrmoke} b) is 11.4~GHz/T.  
The Gilbert damping parameter is found to be roughly independent of the applied field and has an average value of $0.019$. By fitting $f_{res}$ versus applied field the 
effective anisotropy of the film can 
also be evaluated. 
The extracted value of value of 4.5~T obtained from fitting is in excellent 
agreement with the value obtained from the SQUID measurement. 
This, combined with a single peak in the FFT of the spectra, indicate that the uniform precession mode is the dominant contribution to the TR-MOKE signal.
The values of $f_{res}$ and $\alpha$ obtained for this film also compare well to those in the literature~\cite{MWS11}.
Furthermore, they demonstrate the potential use of high anisotropy alloys as microwave oscillators beyond 100~GHz. 

%%%%%%%%%%%%%%%%%%%%%%%%%%%%%%%%%%%%%%%%%%%%%%%%%%%%%%%%%%%%%%%%%%%%%%%%%%%%%%%%
\section{High-field EHE}

As previously mentioned, the anisotropy field of the Mn--Co--Ga film is beyond the 
accessible range of both the SQUID and the VSM (maximum field of 6.5~T and 14~T, respectively).
For the Mn--Co--Ga sample, it was also not possible to obtain any ferromagnetic 
resonance data from the TR-MOKE which is most likely due to the inability to obtain a large enough canting angle for polar MOKE detection.
We therefore determine the anisotropy field using EHE as a detection method.

%%%%%%%%%%%%%%%%%%%%%%%%%%%%%%%%%%%%%%%%%%%%%%%%%%%%%%%%%%%%%%%%%%%%%%%%%%%%%%%%
\begin{figure}[htb]
\centering
\includegraphics[width=7cm]{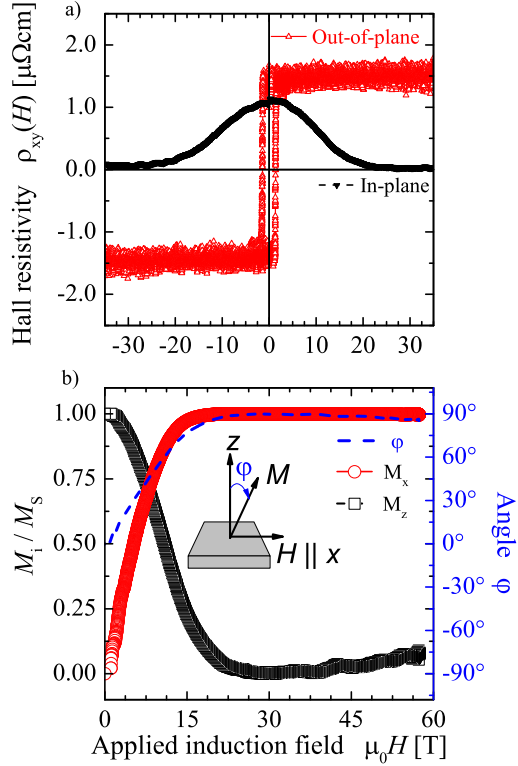}
\caption{ a) EHE response of the Mn--Co--Ga film, measured with in--plane 
             and out--of--plane fields up to 35~T. 
          b) reconstructed in--plane magnetisation curve ($M_x/M_S$) 
             and angle of magnetisation ($\phi$) as a function of applied field up to approximately 60~T. }
\label{fig:coehe}
\end{figure}
%%%%%%%%%%%%%%%%%%%%%%%%%%%%%%%%%%%%%%%%%%%%%%%%%%%%%%%%%%%%%%%%%%%%%%%%%%%%%%%%

Figure~\ref{fig:coehe}~a) shows the EHE curves for both in--plane and out--of--plane
applied fields up to 35~T. The EHE response with the field applied out--of--plane (open red triangles) is, again, identical to 
that obtained from SQUID measurements with sharp switching of the magnetisation visible at 
approximately 1~T. When the magnetic field is applied in the plane of the 
sample (closed black triangles) the magnetisation gradually cants to the plane of the 
film. 
Since the EHE signal is only sensitive to the out--of--plane component of magnetisation, $M_z$, the in--plane component 
of the magnetisation, $M_x$, must be reconstructed.
This is achieved using the transform, $\sin(\cos^{-1}(M_z))$
because the projection of the 
magnetisation along $z$ is simply $\cos(\phi)$, where $\phi$ is the angle of 
the magnetisation to the film normal (see Figure~\ref{fig:coehe} b) inset and Figure~\ref{fig:structure}b)).

Figure~\ref{fig:coehe}~b) shows the both the $M_z$ component (open red circles), the reconstructed in--plane 
$M_x$ component (open black squares), and the canting angle, $\phi$ (blue dashed line). Here, both $M_x$ and $M_z$ 
have been normalised to the maximum. 
It can be seen from the figure that the $M_x$ component is almost totally saturated at 12~T and is completely saturated at 18~T.
The anisotropy energy is obtained by 
evaluating the integral $\int_0^{M_S} \mu_0H_{x}.dM_x$ between zero and saturation, 
i.e the area enclosed by the reconstructed $M_x$ curve and the $y$-axis, 
and multiplying this by $M_S$. 
Beyond saturation the $M_z$ component of magnetisation begins to increase. This is attributed to a small offset between the applied field direction and the film plane. Although the change is rather large in $M_z$ it is corresponds to a misalignment of less than 5$^\circ$.
%By normalising the in--plane curve and taking the area between the curve and the magnetisation axis 
%we can determine the anisotropy energy, akin to what was performed in the case of Mn--Ga.

From the integration of the area between the reconstructed in-plane and out-of-plane loops a value of $K_{eff}$ of 1.23~MJ/m$^3$ is obtained. Although the TR-MOKE results were inconclusive with the Mn--Co--Ga sample, the given value of anisotropy field and saturation magnetization yield precession frequencies of order 350~GHz. 

%%%%%%%%%%%%%%%%%%%%%%%%%%%%%%%%%%%%%%%%%%%%%%%%%%%%%%%%%%%%%%%%%%%%%%%%%%%%%%%%
\section{Conclusion}

We have investigated static and dynamic magnetic properties of Mn$_{1.6}$Ga and Mn$_{2.6}$Co$_{0.3}$Ga$_{1.1}$ films. 
As expected, these materials possess a low saturation magnetisation and high uniaxial anisotropy, usually beyond the accessible field range available in typical SQUID magnetometers. Consequently, standard magnetometry was combined with both time 
resolved magneto-optical Kerr effect and extraordinary Hall effect in high 
magnetic fields to ascertain exact values for the magnetic anisotropy of both types of thin films. 

For the pure Mn$_{1.6}$Ga alloy, we find a magnetic anisotropy energy of 
0.81~MJ/m$^3$, with a saturation magnetisation of 0.36~MA/m. The partial 
substitution of Mn by Co increases the effective anisotropy energy to 
1.23~MJ/m$^3$ and decreases the saturation magnetisation of 0.2~MA/m.

Time-resolved magneto-optical Kerr effect and extraordinary Hall 
effect were utilised to directly probe the anisotropy fields of both materials.
The measured values are 4.5~T and 18~T for Mn$_{1.6}$Ga and 
Mn$_{2.6}$Co$_{0.3}$Ga$_{1.1}$, respectively.

In summary, we have shown that both time-resolved magneto-optical Kerr-effect and the extraordinary Hall effect in high magnetic fields are extremely useful techniques in determining the anisotropy energy for materials which cannot be saturated in standard magnetometers such as SQUID. Such indirect techniques can be readily applied to technologically relevant high anistropy materials for spin-transfer-torque applications.

%%%%%%%%%%%%%%%%%%%%%%%%%%%%%%%%%%%%%%%%%%%%%%%%%%%%%%%%%%%%%%%%%%%%%%%%%%%%%%%%
\ack
C.F. would like to acknowledge Dr. Marc Uhlarz for fruitful discussions.
Financial support by DfG-JST is gratefully acknowledged (projects P~1.3-A and 
P~2.1-A of the Research Unit FOR 1464 {\it ASPIMATT}). Part of this work was funded by EuroMagNET under EU contract no. 228043.

%%%%%%%%%%%%%%%%%%%%%%%%%%%%%%%%%%%%%%%%%%%%%%%%%%%%%%%%%%%%%%%%%%%%%%%%%%%%%%%%
\newpage
\bigskip
\bibliography{MnCoGa}
\bibliographystyle{unsrt}

\end{document}